\documentstyle[12pt]{article}

\advance \textheight by \topskip
\begin{document}

\begin{titlepage}
\hbox to \hsize{\hfil gr-qc/9312004}
\hbox to \hsize{\hfil IHEP 93--121}
\hbox to \hsize{\hfil October, 1993}
\vfill
\large \bf
\begin{center}
CONFORMAL KILLING VECTORS IN \\
FIVE-DIMENSIONAL SPACE.

\end{center}
\vskip 1cm
\normalsize
\begin{center}
{\bf G. L. Rcheulishvili\footnote{E--mail:
rcheul@mx.ihep.su}}\\
{\small Institute for High Energy Physics, 142284 Protvino, Moscow Region,
Russia}
\end{center}
\vskip 2.cm
\begin{abstract}
\noindent
The solutions of generalized Killing equation have been obtained for
line element with initial $t^2 \oplus so(3)$ symmetry. The coefficients
of the metric $g$ corresponding to these vector fields are written down.
\end{abstract}
\vfill
\end{titlepage}

\section{Introduction}

Conformal Killing vectors for  line element (\ref{0}) are being studies in the
paper. In the differential geometry
the invariance of some tensor field $T$ describes the conservation laws in the
presence of the Killing vectors. The conservation laws for traceless
tensors $T$ are connected with the conformal Killing vectors.
Such fields are  characterized by conserved quantities,
whose number is equal to that of conformal Killing vectors [1].
 In classical works [2] the conformal Killing vectors  in the Riemannian
space were investigated from the viewpoint of general principles. The
role played by the vector fields in classifying  the solutions
of the gravitation equations [3],
as well as in other applications [4] is well known.

In works [5] concrete series of isometric Killing vectors, derived from
four-dimensional
line element under its certain parameterization, were studied. A similar
problem was solved in refs. [6,7],
where isometry group generators in five-dimensional space were obtained
for some line
elements. In Kaluza-Klein type theory the five-dimensional space is
 distinguished due to the fact
that out of all mutli-dimensional spaces, just in this space the total
manifold $M^5$ together with  Kaluza-Klein type manifold $M^4 \times S^1$
satisfy  classical Einstein equations [8].
Note, that the solution of the system of equations (\ref{2}) for conformal
Killing vectors
brings us to more clear understanding of the nature of the restrictions
on $g_{nn}$ for isometry groups
obtained in refs. [6,7].

So, we study a five-dimensional space with the metric
\begin{equation}
ds^2 = g_{11}(r) dt^2 - g_{22}(r) dr^2 - 4 g_{33}(r) (\sigma _x^2 +
 \sigma _y^2) -  4  g_{55}(r) \sigma _z^2, \label{0}
\end{equation}
written down in the Cartan's basis. Here  $g_{11}$, $g_{22}$, $g_{33}$,
and $g_{55}$
 are the functions  of the variable $r$;
$d\sigma _x = 2 \sigma _y \land \sigma _z$
( cyclic). Having introduced the variable
\[
z = \int \limits _{}^{r} \sqrt{g_{22}} dr
\]
we rewrite the line element in the form convenient for further studies:
\begin{eqnarray}
ds^2 = g_{11}(z) dt^2&-&dz^2 - g_{33}(z) d \vartheta ^2
\nonumber \\[-1ex]
 \label{1}\\[-1ex]
- (\, g_{33} (z) \sin ^2 \vartheta + g_{55}(z) \cos ^2 \vartheta \, ) d
\varphi ^2
&-&2g_{55}(z) \cos \vartheta  d \varphi d \psi - g_{55}(z) d\psi ^2.
\nonumber
\end{eqnarray}

In the Riemannian  manifold, with the connection that is consistent with the
metric $\tilde g$, the generalized Killing equation for the
component $\xi ^k$ of the vector
 $\zeta = \xi ^k \partial _k$ looks like:
\begin{equation}
\xi ^k{\partial }_k \tilde g_{ij}+ \tilde g_{ik}{\partial }_j \xi ^k+
 \tilde g_{kj}{\partial }_i \xi ^k= \Phi \tilde g_{ij} .\label{2}
\end{equation}
Here  ${\partial }_1 =
 \frac{{\partial }}{{\partial }t}$, ${\partial }_2 = \frac{{\partial }}
{{\partial }z}$, ${\partial }_3 = \frac{{\partial }}{{\partial }\vartheta }$,
 ${\partial }_4 = \frac{\partial }{\partial \varphi }$,  ${\partial }_5 =
\frac{\partial }{\partial \psi }$.

Our goal is to obtain all series of vectors, that give the conformal
motions in space (\ref{2}).
At the same time, the highest symmetries
(as compared with the initial symmetry $L = t^2 \oplus so(3)$ of the
line element)
take place under certain restrictions imposed on
$g_{nn}(z)$. In most cases these restrictions are
 explicitely solved w.r.t. $g_{nn}$,
which gives us a possibility to write down suitable $g_{nn}$, that ensure
the existence of the given series of vectors.

\section{Solution of the System}

Let us slightly transform the equations for
$\tilde g_{44}$  and  $\tilde g_{45}$ in system  (\ref{2})
so that in the R.H.S. of the equation, that corresponds to the
component $\tilde g_{44}$, we would have
$\frac{1}{2}\Phi $, and the R.H.S. of the equation for the component
$\tilde g_{45}$ would be equal to zero.
Taking into account this modification, the equations in system (\ref{2}) will
 be numerated by the indices $\tilde g_{ij}$: $(i,j)$.
Besides we shall introduce  notations for the additional
equations obtained by subtracting from each other the
 equations, containing in the R.H.S. $\Phi $  (i.e. the equations of the
type $(n,n)$ ): $\tilde{(i,k)} = (i,i) -(k,k) $.

One of the possible ways to solve system (\ref{2}) is to apply the
integrability conditions. Let us attempt to
maximally simplify the procedure. By differentiating the
equations with wave over $\partial _{ik}$: $\partial
_{ik}\tilde{(i,n)}$, where $i\ne k,\; n\ne k, \quad i,k,n =
1,2,3,5$ and using the remaining equations $(j,l)$ of system (\ref{2}),
we shall obtain pairs of equations of the third order for the quantity
$\tilde {\xi }^k, \quad k = 1,2,3,5$:
\begin{eqnarray}
& &\partial _{iii}\xi ^2 + g_{ii}(z) R_i(z) \partial _i \xi ^2 =
0 , \nonumber\\
& &\partial _{iii}\xi ^2 + g_{ii}(z) \hat R_i(z) \partial _i \xi ^2 =
0 ,   \qquad i = 1,3,5; \label{3}
\end{eqnarray}
and
\begin{eqnarray}
& &\partial _{222} \tilde {\xi }^i + 3 F_i \partial _{22} \tilde
{\xi }^i + \{
(F'_i + 2 F_i^2) + \hat{\gamma } R_i (z) \} \partial _2 \tilde {\xi }^i =
0, \nonumber \\
& &\partial _{222} \tilde {\xi }^i + 3 F_i \partial _{22} \tilde
{\xi }^i + \{
(F'_i + 2 F_i^2) + \hat{\gamma } \hat R_i (z) \} \partial _2
\tilde {\xi }^i
= 0,  \qquad i = 1,3,5;  \label{4} \\
& &\qquad  \tilde {\xi }^k = \xi ^k,  \quad  k =1,3; \qquad \tilde {\xi }^5 =
\xi ^5 +
\cos \vartheta \xi ^4.\nonumber
\end{eqnarray}
Here $\hat{\gamma } = 1$ for $i = 1$  and  $\hat{\gamma } = -1$
for $i =3,5$;   $F_i =
\frac{g'_{ii}}{2g_{ii}}$, $i = 1,3,5$. Here and in what follows the
prime will denote the derivative
over the variable $z$. The quantities $R_i$ and $\hat R_i$ have the
following form:
\begin{eqnarray*}
& &R_1(z) = (F_1 - F_3)' + F_3 (F_1 - F_3),  \\
& &\hat R_1(z) = (F_1 - F_5)' + F_5 (F_1 - F_5);
\end{eqnarray*}
\begin{eqnarray}
& &R_3(z) = (F_1 - F_3)' + F_1 (F_1 - F_3), \nonumber \\
& &\hat R_3(z) =\frac{1}{2} \{  (F_5 - F_3)' + F_5 (F_5 - F_3) +
 \frac{1}{g_{33}} (1 - \frac{g_{55}}{2 g_{33}}) \}; \label{6}
\end{eqnarray}
\begin{eqnarray}
& &R_5(z) = (F_1 - F_5)' + F_1 (F_1 - F_5), \nonumber \\
& &\hat R_5(z) = (F_3 - F_5)' + F_3 (F_3 - F_5) +
\frac{g_{55}}{4 g_{33}^2}. \label{7}
\end{eqnarray}
{}From formulae (\ref{3}) and (\ref{4}) we have:
\begin{equation}
(R_i - \hat R_i) \partial _i \xi ^2 =(R_i - \hat R_i) \partial
_2 \tilde {\xi }^i = 0, \qquad i =1,3,5. \label{8}
\end{equation}
Besides from the combination of equations (\ref{3}) and (\ref{4})
there follows the restriction for $\hat R_i$. The function $\hat R_i$
should satisfy the condition
\begin{equation}
\hat R_i = \frac{{\rm const}}{g_{ii}} ,\label{9}
\end{equation}
where const may also be zero.

The following remark should be made: from system (\ref{2}) one can easily
find out, that the equality to zero of the derivative
  $\partial _2 \xi ^i = 0$ (or $\partial _i \xi ^2 = 0$),
where $i$ takes one of three values
$i = 3,4,5$, entails the equality to zero of other similar quantities.
Hence we have
\begin{eqnarray}
& & \partial _2 \xi ^i = 0  \;  (\partial _i \xi ^2 = 0) \;
\Rightarrow \;  \{ \partial _i \xi ^2 = 0 \; ( \partial
_2 \xi ^i = 0 ),  \nonumber \\[-1ex]
\label{10}\\ [-1ex]
& & \partial _j \xi ^2 = 0, \;  \partial _2 \xi ^j = 0, \;
\partial _k \xi ^2 = 0, \; \partial _2 \xi ^k = 0 \}; \qquad
i,j,k = 3,4,5 .\nonumber
\end{eqnarray}

This observation greatly simplifies the problem, only
demanding simultaneous dependence of
$\xi ^2$ on $\theta , \varphi $ and
$\psi $. Formula (\ref{8}) shows, that in the case of
$\xi ^2 = \xi ^2(\theta , \varphi , \psi )$
both equations $R_3 = \hat R_3$ and
$R_5 = \hat R_5$ should only be fulfilled together.

First we shall look for the explicit dependence of the components
$\xi ^2$, $\xi ^3$,
$\xi ^4$  and  $\xi ^5$ on the variables $z, \vartheta , \varphi
$ and
$\psi $, i.e. we shall solve the equations $(i,k)$,
$\tilde{(i,k)}$, where $i,k \ne 1$. Formulae (\ref{8}),
(\ref{9}) and (\ref{10}) make it clear that a nontrivial solution is
posssible only in the following
cases:
\begin{equation}
{\rm either} \quad  {\rm I}. \qquad \qquad  \partial _i \xi ^2 = \partial _2
\xi ^i = 0,
\qquad \qquad i =3, 4 ,5;  \qquad  \qquad \label{11}
\end{equation}
\begin{eqnarray}
{\rm or}   \qquad   {\rm II}. \quad   \qquad
 & &R_3 = \hat R_3, \quad R_5 = \hat R_5, \qquad
{\rm and}\label{100} \\
a). \quad & &R_i = 0, \qquad i =3, 5; \qquad {\rm or}\nonumber \\
b). \quad & &R_i = \frac{\rho _{ii}}{g_{ii}}, \quad i =3,5,
\qquad  \rho _{ii} = {\rm const} \ne 0. \qquad  \qquad   \label{13}
\end{eqnarray}

In the simplest case I, when $\xi ^2 = \xi ^2(t,z)$ and as a consequence
equalities (\ref{100}) are not fulfilled,
conditions (\ref{11})  bring us to three different series $V_a$,
$V_b$ and  $V_c$  for the components $\xi ^2 \div   \xi ^5$. Let us
write down two components $\xi ^2$ and $\xi ^3$ for each of these series.

\noindent $V_a: \quad \{ F_3 - F_5 \ne 0, \;  R_3 \ne \hat R_3,
 \;  R_5 \ne \hat R_5 \}$
\[
\xi ^2 = 0,  \quad \xi ^3 = \xi ^3_1(t) \sin {\varphi } + \xi ^3_2(t) \cos
{\varphi }.
\]
$V_b: \quad \{ g_{33} = \tau _3 g_{55}, \; \tau _3 =
{\rm const} \ne 1 \}$
\[
\xi ^2 = \sqrt {g_{33}} \xi ^2_1(t), \quad  \xi ^i = \xi ^i_{V_a};
\qquad i = 3,4,5.
\]
$V_c: \quad \{ g_{33} = g_{55}, \;  R_3 \ne \hat R_3 \}$
\[
\xi ^2 = \sqrt {g_{33}} \xi ^2_1(t), \quad  \xi ^3 = \xi ^3_{V_a}
+ \xi ^3_3(t) \sin {\psi } - \xi ^3_4(t) \cos {\psi }.
\]

As for the second possibility, described with formula (\ref{100}),
we must consider four different cases: i) $R_3 = R_5 = 0, \quad $
ii) $R_3 = 0, \; R_5 = \frac{\rho
_{55}}{g_{55}}, \quad  $  iii) $R_3 = \frac{\rho _{33}}{g_{33}}, \;
R_5 = 0 \quad  $ and  iv)  $R_3 = \frac{\rho _{33}}{g_{33}}, \; R_5 =
\frac{\rho _{55}}{g_{55}}$.

The key component here is $\xi ^2$. The dependence on $\vartheta $ and
$\psi $  is fixed by equations (\ref{3}).
It is convenient to derive the dependence $\xi ^2$ on the variable z
from the analysis of equations $\tilde{(2,3)}$,  $\tilde{(2,5)}$ and
$\tilde{(3,5)}$.
Using $(2,3)$ ¨ $(2,5)$, one can bring them to the form
\begin{eqnarray}
& &\partial _{22} \xi ^2 - \partial _2 (F_3 \xi ^2) + \frac{1}{g_{33}}
\partial _{33} \xi ^2 = 0, \nonumber \\
& &\partial _{22} \xi ^2 - \partial _2 (F_5 \xi ^2) + \frac{1}{g_{55}}
\partial _{55} \xi ^2 = 0, \label{17} \\
& &\partial _2 \{(F_3 - F_5) \xi ^2 \}  - \frac{1}{g_{33}} \partial _{33}
\xi ^2  +  \frac{1}{g_{55}} \partial _{55} \xi ^2 = 0. \nonumber
\end{eqnarray}

However it can easily be seen that these equations for
conditions i), ii) and iii)
are consistent only in the trivial case  $\xi ^2 \ne \xi ^2 (\vartheta ,
\varphi ,\psi )$ and result in the series $V_a$, $V_b$, $V_c$.

In case iv), $R_3 = \frac{\rho _{33}}{g_{33}}$ and
$R_5 = \frac{\rho _{55}}{g_{55}}$, let $\xi ^2$ be presented in the
following way
\begin{equation}
\xi ^2 (t,z,\vartheta ,\varphi ,\psi ) = \sum _{i,k = 0}^{2}  \xi ^{2(ik)}
(t,z,\varphi ) m_i(\vartheta ) f_k(\psi ). \label{18}
\end{equation}
The explicit dependence on the variables $\vartheta $ and  $\psi $ is given
by the functions
\begin{equation}
m_0 = 1,  \quad
m_1 = \left\{
\begin{array}{l}
\sin {\sqrt{\rho _{33}} \vartheta } \\
{\rm sh}{\sqrt{|\rho _{33}|} \vartheta },
\end{array}\right.
\; m_2 = \left\{
\begin{array}{l}
\cos {\sqrt{\rho _{33}} \vartheta } \\
{\rm ch}{\sqrt{|\rho _{33}|} \vartheta }
\end{array}\right.
 \quad {\rm for} \quad
\begin{array}{l}
\rho _{33} > 0\\
\rho _{33} < 0
\end{array}
\label{19}
\end{equation}
and by similar $f_i(\psi )$ but with the replacement
$\rho _{33} \to  \rho _{55}$
and  $\vartheta   \to   \psi  $.

Before starting to search for the dependence $\xi ^{2(ik)}(t,z,\varphi )$
on $z$,  we shall  put down conditions
(\ref{13}) in their explicit form
\begin{eqnarray*}
& &(F_5 - F_3)' + F_5 (F_5 - F_3) + \frac{1 - 2\rho
_{33}}{g_{33}} - \frac{g_{55}}{2g_{33}^2} = 0, \\
& &(F_3 - F_5)' + F_3 (F_3 - F_5) - \frac{\rho _{55}}{g_{55}} +
 \frac{g_{55}}{4g_{33}^2} = 0; \\
& &(F_3 - F_5)^2 + \frac{1 - 2\rho _{33}}{g_{33}} - \frac{\rho _{55}}{g_{55}}
 - \frac{g_{55}}{4g_{33}^2} = 0.
\end{eqnarray*}
The third equation, that is a consequence of the first two equations, is
given here for conveniency.
Equations (\ref{17}) are analyzed separately for a)  $F_3 - F_5 \neq  0$  ¨
b)  $F_3 - F_5 = 0$.

At a) $F_3 - F_5 \neq  0$  equations (\ref{17}) turn out to be inconsistent
for the nonzero values of  $\xi ^{2(k0)}(t,z,\varphi )$
 and  $\xi ^{2(0k)}(t,z,\varphi )$, $k =0,1,2$,  no matter what values
 $\rho _{33}$ and  $\rho _{55}$ take. Only for $\rho _{33} = \rho
_{55} = 1/4$, $\xi ^{2(\mu
\nu )}(t,z,\varphi ) \; (\mu ,\nu =1,2)$ in (\ref{18})
has a nontrivial solution. The function $\sigma (z) \ne {\rm const}$:
\begin{equation}
g_{33} = \sigma (z) g_{55}, \qquad \sigma '(z) \ne 0 \label{21}
\end{equation}
 is introduced to make our further description more
convenient.
Note, that  $F_3 -F_5 = \frac{\sigma '(z)}{2 \sigma (z)}$. In the
terms of the function
$\sigma (z)$ the solution of equation  (\ref{17})
for $\xi ^{2(\mu \nu )}(t,z,\varphi )$, alongside with the expressions
for $g_{nn}$
 satisfying restrictions (\ref{100}),
has a simple form
\begin{equation}
\xi ^{2(\mu \nu )} = A^{\mu \nu }(t,\varphi ) \frac{\sigma
}{\sigma '} \sqrt {(\sigma -1)/\alpha }, \qquad \alpha = {\rm
sign}(\sigma - 1) \label{22}
\end{equation}
and
\begin{equation}
g_{11} = {\rm const}\frac{\sigma ^2 (\sigma  - 1)}{(\sigma ')^2
\alpha }, \qquad  g_{33} = \frac{\sigma (\sigma - 1)^2}{(\sigma
')^2}, \qquad  g_{55} = \frac{(\sigma - 1)^2}{(\sigma ')^2}. \label{23}
\end{equation}

In the second case, b) $F_3 - F_5 = 0$, only $\xi ^{2(\mu \nu )}$
 ($\mu \nu  = 1,2$)
and  $\xi ^{2(00)}$  are nontrivial solutions of the equations and only under
equality of $g_{33} = g_{55}$
and definite values of the constants $\rho _{ii}$: $\rho _{33} = \rho _{55}
= 1/4$.
Elementary calculations yield
\begin{eqnarray}
& &\xi ^2(t,z,\vartheta ,\varphi ,\psi ) = \sqrt {g_{33}}\{ \xi
^{2(\mu \nu \lambda )} (t,\varphi ) n_{\lambda }(\chi _3) m_{\mu
} (\vartheta ) f_{\nu } (\psi ) + \nonumber \\
& &B(t,\varphi ) + G(t,\varphi )
\chi _3 \}, \quad  \qquad \qquad \qquad \qquad  \mu ,\nu ,\lambda  = 1,2.
\label{24}
\end{eqnarray}
Here and in what follows $\,  \chi _3 = \int \limits _{}^{z} \sqrt {\frac{1}
{g_{33}}}dz, \;
n_1(\chi _3) = {\rm sh}\frac{1}{2}\chi _3$ ¨ $ n_2(\chi _3) =
 {\rm ch}\frac{1}{2}\chi _3$.

For both cases a) and b) we obtain the final dependence of the
components $\xi ^2 \div  \xi ^5$ on
the variables $z,\vartheta ,\varphi $ and $\psi $   applying equations
$(2,2) \div   (5,5)$  of system (\ref{2})
to equations (\ref{22}) and (\ref{24}).
Similar to the case $V_a \div  V_c$, we shall write down only the second
and third components:

\noindent $V_f: \quad \{ F_3 - F_5 \ne 0, \;  R_3 = \hat R_3,
 \;  R_5 = \hat R_5 \}$
\begin{eqnarray*}
& &\xi ^2 = \frac{\sigma }{\sigma '} \sqrt {(\sigma  -1)/\alpha
} N^m(t) P_m(\vartheta ,\varphi ,\psi ),  \\
& &\xi ^3 = \frac{2}{\alpha } \sqrt {\frac{\alpha }{\sigma  -1}} N^m \partial
_3P_m + \xi ^3_{V_a}; \qquad   \qquad  m = 8,9,10,11.
\end{eqnarray*}
$V_g: \quad \{ g_{33} = g_{55}, \; R_3 = \hat R_3\}$
\begin{eqnarray*}
& &\xi ^2 = \sqrt{g_{33}} \{ N^{\mu m}(t) n_{\mu }(\chi _3)
P_m(\vartheta ,\varphi ,\psi ) + M(t)\}, \\
& &\xi ^3 = -2\{N^{1m} n_2 + N^{2m} n_1\}\partial _3P_m + \xi
^3_{V_c} ; \qquad \mu =1,2.
\end{eqnarray*}
 Here  $P_m (\vartheta ,\varphi ,\psi )$ signify the functions:
\begin{eqnarray*}
& &P_8 = \cos \frac{\varphi  -  \psi }{2}\sin {\vartheta /2}, \qquad
P_9 = - \sin \frac{\varphi  - \psi }{2}\sin {\vartheta /2},\\
& &P_{10} = \cos \frac{\varphi  + \psi }{2}\cos {\vartheta /2}, \qquad
P_{11} = \sin \frac{\varphi  + \psi }{2}\cos {\vartheta /2}.
\end{eqnarray*}
Note that  $(\partial _{34} + \frac{1}{2} \sin \vartheta  \partial _5 +
\cos \vartheta  \partial _{35}) P_m = 0$.

For $V_a \div   V_f$ the dependence on $t$ is given by equations
(\ref{3}). The final form of the series we get applying equations
$(1,2) \div   (1,5)$  and  $\tilde{(1,2)} \div  \tilde {(1,5)}$
to the sets  $V_a \div   V_f$. As in the case with formulae (\ref{11})
$\div $ (\ref{13})  equations (\ref{3}), (\ref{4}), (\ref{8})  and
(\ref{9}) show, that for this  one should take into consideration
only two possibilities:
\begin{equation}
{\rm either} \quad  {\rm I}.  \qquad \qquad  \partial _1 \xi ^2 =
\partial _2 \xi ^1 = 0,
\qquad \qquad i =3, 4 ,5;  \qquad  \qquad  \label{33}
\end{equation}
\begin{eqnarray}
{\rm or}   \qquad   {\rm II}. \qquad   \qquad  & &R_1 = \hat R_1, \qquad
\qquad  {\rm and} \label{30} \\
a). \quad & &R_1 = 0, \quad  \qquad  \qquad {\rm or}\label{32} \\
b). \quad & &R_1 = \frac{\rho _{11}}{g_{11}}, \;  \qquad
\qquad  \rho _{11} = {\rm const} \ne 0. \qquad  \qquad  \label{31}
\end{eqnarray}
Let us successively use these restrictions for the vectors $V_a \div  V_f$.
Together with the final form of the series we shall write down
  corresponding $g_{nn}$,  that are solutions
 of restrictions (\ref{13}), (\ref{32})  ¨  (\ref{31}).

\section{Results}

First it should be noted that the series with minimal number of vectors,
five and seven,
are derived from  $V_a$,  $V_b$  and  $V_c$ under condition (\ref{33})
 ($V_a$ leads to $V_5$ for all formulae (\ref{33}) $\div $ (\ref{31})):

\noindent $V_5: \quad \{ R_i \ne \hat R_i, \; i = 3,5;
(g_{33} = \tau _3 g_{55}, \, R_1 \ne  0, \,
 R_1 \ne  \frac{\rho _{11}}{g_{11}}, \; {\rm or} \, F_3 - F_5
\ne 0)   \}$
\begin{eqnarray*}
\xi _5^1= D^7, \; \xi _5^2= 0,\, & &\xi _5^3= D^1\sin
\varphi  + D^2\cos \varphi , \; \xi _5^4={\rm ctg} \vartheta (D^1\cos
\varphi - \\
 D^2 \sin \varphi ) - D^5,\, & &\xi _5^5= -\frac{1}{\sin \vartheta }
(D^1\cos \varphi  - D^2 \sin \varphi ) + D^6.
\end{eqnarray*}
$V^0_7: \quad \{ g_{33} = g_{55}, \;  R_1 \ne  0, \;  R_1 \ne
\frac{\rho _{11}}{g_{11}},  \; R_3 \ne \frac{1}{4g_{33}} \}$
\begin{eqnarray*}
\xi _7^1 = D^7, \;  \xi _7^2 = 0, & &\xi _7^3 = - D^3
\cos \psi  +  D^4\sin \psi  +  \xi _5^3, \; \xi _7^4 =
- \frac{1}{\sin \vartheta }(D^3\sin \psi +  \\
D^4\cos \psi ) + \xi _5^4, & &\xi _7^5 = {\rm ctg}\vartheta (D^3\sin
\psi  + D^4\cos \psi ) + \xi _5^5.
\end{eqnarray*}
In both cases  $\Phi  =  0$, i.¥.  $V_5$ and  $V^0_7$ are generators
of isometry groups.
Here and in what follows $D^{mk}=$ const, and basis vectors $\zeta _i$
in the series are built from the components   $\xi ^k$
in the following way
\[
\zeta _i = \frac{\partial }{\partial D^i} \xi ^k (\sum _m^{}D^m)
\partial _k.
\]
Two sets $V^i_7$  ¨ $V^i_9$, $i = 1,2,3$, three series in each,
consisting of seven and nine vectors,
are obtained from $V_b$  ¨ $V_c$, respectively, under conditions (\ref{32})
and (\ref{31}).  The first two components are equal at fixed $i$ for
$V^i_7$ ¨ $V^i_9$, while three remaining components correspond to $V_5$
for  $V^i_7$   and  $V^0_7$   for  $V^i_9$.

\noindent $V^1_7 (V^1_9): \qquad  \{g_{11} = \tau _1 g_{33},
 \; \tau _1 = {\rm const}  \}$
\[
\xi ^1 = \frac{\chi _3}{\tau _1} D^{20} + D^7, \qquad \xi ^2 =
\sqrt{g_{33}} (D^{20}t + D^{21});
\]
$V^2_7 (V^2_9): \qquad  \{g_{11} = g_{33} b {\rm exp}(2\hat \delta \chi
_3), \; b,\hat \delta  ={\rm const},  (\hat \delta ^2\ne \frac{1}{4} \;
{\rm for} V_9^2)\}$
\begin{eqnarray*}
& &\xi ^1 = - (\frac{1}{2 b \hat \delta }{\rm exp}(-2\hat \delta \chi
_3) + \frac{\hat \delta }{2} t^2) D^{20} - \hat \delta  tD^{21} +
D^7, \\
& &\xi ^2 = \sqrt{g_{33}} (D^{20} t + D^{21});
\end{eqnarray*}
$V^3_7 (V^3_9): \qquad  \{(F_1 - F_3)' + F_3 (F_1 - F_3) - \frac{\rho
_{ii}}{g_{11}} = 0\}$
\begin{eqnarray*}
& &\xi ^1 = \pm \sqrt {\frac{g_{33}}{|\rho _{11}|}} (F_1 - F_3)
(D^{20}l_2(t) \mp D^{21}l_1(t)) + D^7, \\
& &\xi ^2 = \sqrt{g_{33}} (D^{20}l_1(t) + D^{21}l_2(t)).
\end{eqnarray*}
For other three components:
\begin{eqnarray*}
\xi ^i = \xi ^i_{V_5}, \quad {\rm for} & &V^k_7 \; (g_{33} = \tau _3
g_{55},\;  \tau _3 \ne 1); \\
\xi ^i = \xi ^i_{V^0_7}, \quad {\rm for} & & V^k_9 \; (g_{33} = g_{55});
 \qquad i =3,4,5, \qquad k = 1,2,3.
\end{eqnarray*}
The functions $l_{\mu }$ are given by the formula similar
to (\ref{19}) but
with the replacements  $\vartheta \rightarrow  t$  and  $\rho _{33} \to  \rho
_{11}$. Two signs in $\xi ^1$  for $V^3_7 (V^3_9)$  are related
to $\rho _{11} > 0$  and $\rho _{11} < 0$, respectively.
In these series we have: $R_3 \ne  \hat R_3$  ¨
$R_5 \ne  \hat R_5$. The function  $\Phi $  for the series  $V^i_7
(V^i_9), \;  i = 1,2,3,$ looks like $\Phi  = 2 F_3 \xi ^2$.

The other series are connected with the sets  $V_f$  ¨  $V_g$,
whose third components depend on the variable $z$.
Since  equation (\ref{30})
for $V_f$  is not fulfilled, the only possible condition (\ref{33})
 applied to $V_f$ leads to the series, containing  nine vectors:

\noindent $V^4_9: \qquad    \{g_{nn} \;{\rm are \; given \; by
\;  formula \;} \;
(\ref{23})\}$
\begin{eqnarray*}
& &\xi ^1 = D^7, \qquad  \xi ^2 = \frac{\sigma
}{\sigma '}  \sqrt {(\sigma  -1)/\alpha } D^m P_m, \\
& &\xi ^3 = \frac{2}{\alpha } \sqrt {\frac{\alpha }{\sigma  -1}} D^m \partial
_3P_m +  \xi ^3_{V_5}, \\
& & \xi ^4 = - \frac{4}{\alpha \sin \vartheta } \sqrt {\frac{\alpha }
{\sigma  -1}} D^m \partial
_{35}P_m + \xi ^4_{V_5}, \\
& &\xi ^5 = \frac{2}{\alpha } \sqrt {\frac{\alpha }{\sigma  -1}}
D^m\{(2 - \sigma ) \partial
_5P_m + 2 \cot \vartheta \partial _{35}P_m\} + \xi ^5_{V_5};\\
& &\qquad  \qquad  \qquad  \Phi  = 2 F_1 \xi ^2; \quad  m =8,9,10,11.
\end{eqnarray*}

It is the set $V_g$, that gives the largest number of series.
It yields a series  consisting of eleven vectors
and three series, containing twenty one conformal Killing vectors,
i.e. the maximum possible number.
The series consisting of eleven vectors $V_{11}$ are obtained
from $V_g$  under condition  (\ref{33}):

\noindent $V_{11}: \, \{ g_{33} = g_{55}, \; R_1 \ne 0, R_1
\ne \frac{\rho _{11}}{g_{11}}, \;  (F_1 - F_3)' + F_1(F_1 -
F_3) - \frac{1}{4 g_{33}} = 0\}$
\begin{eqnarray*}
& &\xi ^1 = D^7, \;  \xi ^2 = \sqrt{g_{11}} D^m P_m,
\; \xi ^3 = -4(F_1 - F_3)\sqrt{g_{11}} D^m \partial _3P_m + \xi
^3_{V^0_7}, \\
& &\xi ^{4+i} = \frac{8}{\sin \vartheta }(F_1 - F_3)\sqrt{g_{11}}
D^m \partial _{3(5-i)}P_m + \xi ^{4+i}_{V^0_7};\\
& &\qquad    \Phi  = 2 F_1 \xi ^2; \qquad  \qquad  i = 0,1;
\qquad  m = 8,9,10,11.
\end{eqnarray*}

Before we start describing the remaining series, we shall make a small remark.
As is seen from  the condition $R_3 = \hat R_3$, for  $V_g$
we have $F_1 - F_3 \ne  0$. Similar to formula (\ref{21}) this
fact allows us to introduce a nontrivial function of the variable
$z$:   $g_{11} = \sigma _1(z)g_{33}$, $\; \sigma _1'(z) \ne
0$. Conditions (\ref{13}), (\ref{32}) and   (\ref{31}) are
solved w.r.t. $\sigma _1$ and allow one to present the
corresponding $g_{ii}$ in terms of this function.

As a result, applying condition (\ref{31}) to  $V_g$ we obtain two series
with twenty one generalized Killing vectors \vspace{0.5ex}

\noindent $V^1_{21}: \qquad  \{g_{11} = \frac{(\sigma _1)^2
(\sigma _1 - 4 \rho _{11})}{(\sigma _1')^2},
  \quad  g_{33} = g_{55} =  \frac{\sigma _1 (\sigma _1- 4 \rho _{11})}
{(\sigma _1')^2},  \quad   \sigma _1' \ne  0\}$
\begin{eqnarray*}
& &\xi ^1 = \sqrt{\frac{|\rho _{11}|}{\sigma _1}} \{ -2 (D^{m1} l_2(t) \mp
 D^{m2} l_1(t)) P_m + \frac{\sqrt{\sigma _1 - 4 \rho _{11}}}{2
\rho _{11}} (D^{20}l_2 \mp  \\
& &D^{21}l_1)\}+ D^7, \qquad  \qquad   \xi ^2 = \frac{\sqrt{\sigma _1
(\sigma _1 - 4 \rho _{11})}}{(\sigma _1')}
 \{(\sqrt{\sigma _1 - 4 \rho _{11}} D^{m\mu }l_{\mu } + \\
& &\sqrt{\sigma _1} D^{m3}) P_m +  D^{20}l_1 + D^{21}l_2\}, \\
& &\xi ^3 = - 2 \{\sqrt{\sigma _1} D^{m\mu }l_{\mu } +
\sqrt{\sigma _1 - 4 \rho _{11}}D^{m3}\}\partial _3P_m + \xi
^3_{V^0_7},\\
& &\xi ^{4+i} = \frac{4}{\sin \vartheta } \{\sqrt{\sigma _1} D^{m\mu }
l_{\mu } +
 \sqrt{\sigma _1 - 4 \rho _{11}}D^{m3}\}\partial _{3(5-i)}P_m + \xi
^{4+i}_{V^0_7}; \\
& &\Phi  = 2\{(F_3 \sqrt{g_{33}(\sigma _1 -4 \rho _{11})}\beta +
\frac{\sqrt{\sigma _1}}{2}) D^{m\mu }l_{\mu }P_m + \sqrt{g_{11}} F_1
 \beta D^{m3}P_m + \\
& &\sqrt{g_{33}} F_3 \beta  (D^{20} l_1 + D^{21} l_2\};
 \qquad   \qquad  \beta ={\rm sign} \sigma _1'; \qquad  i,\mu = 1,2.
\end{eqnarray*}
Two signs for  $\xi ^1$ are related to $\rho _{11} > 0$ and  $\rho
_{11} < 0$, respectively.

And finally condition (\ref{32}) leads to series $V^2_{21}$:

\noindent $V^2_{21}: \qquad  \{g_{11} = \frac{(\sigma _1)^3}{(\sigma _1')^2},
  \quad  g_{33} = g_{55} =  \frac{(\sigma _1)^2}{(\sigma _1')^2},
  \quad    a = {\rm const} > 0\}$
\begin{eqnarray*}
& &\xi ^1 = - \frac{2}{a} \sqrt{\frac{a}{\sigma _1}} (D^{m2}t +
D^{m1}) P_m - D^{21}(t^2/4 + 1/\sigma _1) - \frac{1}{2} D^{20}t +
D^7, \\
& &\xi ^2 = \frac{\sigma _1}{\sigma _1'}\{\sqrt{\frac{\sigma
_1}{a}}[D^{mr}t^r/r! - (2/a - 2/\sigma _1) D^{m2}] P_m +
  D^{21}t + D^{20}\}, \\
& &\xi ^3 =  - 2 \sqrt{\frac{\sigma
_1}{a}}\{D^{mr}t^r/r! - (2/a + 2/\sigma _1) D^{m2}\} \partial _3P_m +
 \xi ^3_{V^0_7}, \\
& &\xi ^{4+i} = \frac{4}{\sin \vartheta } \sqrt{\frac{\sigma
_1}{a}}\{D^{mr}t^r/r! - (2/a + 2/\sigma _1) D^{m2}\}
\partial _{3(5-i)}P_m + \xi ^{4+i}_{V^0_7},\\
& &\Phi = 2 \frac{\sigma _1}{\sigma _1'}\{\sqrt{\frac{\sigma
_1}{a}}[(F_1(t^2/2 - 2/a -2/\sigma _1) + \frac{4}{\sigma _1} F_3)
D^{m2} + \\
& & F_1(D^{m1}t + D^{m0})] P_m + F_3(D^{21}t +
D^{20}))\}; \qquad  r = 0,1,2.
\end{eqnarray*}
Thus, expressions $V_5 \div  V^2_{21}$  represent all series of
generlazed Killing vectors
for line element (\ref{1}). In accordance with theorem [9] there
follows from the form of the function
$\Phi $  for the series  $V^i_7, \;  V^i_9, \;  i =
1,2,3, \quad  V^4_9$   and  $\; V_{11}$
that these series are generators of trivial conformal groups, and,
consequently,  in a corresponding coordinate system
they coincide with the generators of the isometry groups,
described in [7].

In conclusion the author would like to express his gratitude to
M.V.Saveliev for useful duscussions.

\end{document}